\documentclass[%
 reprint,
showpacs,preprintnumbers,
 amsmath,amssymb,
 aps,
 prl,
 twocolumn,
]{revtex4-1}

\usepackage{graphicx}
\usepackage{subfigure}
\usepackage{dcolumn}
\usepackage{bm}

\begin{document}


\title{Strong Quantum Spin Correlations Observed in Atomic Spin Mixing}

\author{Eva M. Bookjans}
\author{Christopher D. Hamley}
\author{Michael S. Chapman}
\affiliation{School of Physics, Georgia Institute of Technology, Atlanta, Georgia 30332-0430}

\date{\today}

\begin{abstract}
We have observed sub-Poissonian spin correlations generated by collisionally induced spin mixing in a spin-1 Bose-Einstein condensate. We measure a quantum noise reduction of  $-7$ dB ($-10$ dB corrected for detection noise) below the standard quantum limit (SQL) for the corresponding coherent spin states. The spin fluctuations are detected as atom number differences in the spin states using fluorescent imaging that achieves a detection noise floor of 8 atoms per spin component for a probe time of 100 $\mu$s.

\end{abstract}
\maketitle



The study of quantum correlated states including squeezed and entangled states is an active research frontier with important applications in precision measurements, quantum information and fundamental tests of quantum mechanics.  Much of the early research in this area focused on quantum optical systems \cite{Bachor04}, motivated originally by the suggestion that squeezed states could be used in gravity wave detectors to surpass the standard quantum limit \cite{Caves81, Schnab10}. There has also been significant progress in realizing squeezing and other quantum correlated (nonclassical) states in atomic systems, using either non-linear atom-light interactions \cite{*[{For a recent review, see }][] Polzik10}, or more recently,  collisional interactions in ultracold atomic gases \cite{Orzel01,Greiner02,Greiner05,Foelling05,Chuu05, Itah10, Esteve08, Whitlock10,Gross10,Riedel10,Jaskula10,Schmied11}.

In optics, squeezed states of light can be generated using non-linear optical interactions that create quantum correlations between different field modes. An important example is optical four-wave mixing, which is a third-order parametric process employed in the first demonstration of squeezed states of light in the pioneering experiments by Slusher \emph{et al.} 25 years ago \cite{Slusher85}. In spontaneous four-wave mixing, a strong pump field interacting with medium with a $\chi^{(3)}$ non-linearity generates two correlated optical beams known as the signal and idler modes that are exactly correlated in photon number, anticorrelated in phase and exhibit two-mode quadrature squeezing \cite{Walls}.

In ultracold atomic gases, binary \emph{s}-wave collisions between atoms naturally give rise to strong third-order non-linear interactions capable of producing analogous four-wave mixing of atomic matter waves. Indeed, both stimulated and spontaneous atomic four-wave mixing have been observed with colliding condensates \cite{Deng99, Vogels02, Perrin07, Pertot10}, and in the spin dynamics of spinor condensates \cite{Stenger99,Schmaljohann04, Chang04, Chang05, Sadler06, Lett09a}.  Recently, sub-Poissonian correlations were  observed in spontaneous four-wave mixing of two colliding condensates \cite{Jaskula10}, manifest as $-0.5$~dB relative atom number squeezing measured between outgoing modes of opposing momenta in the \emph{s}-wave scattering halo.  The focus of this work is the demonstration of sub-Poissonian spin correlations generated by four-wave spin mixing (4WSM).

In spinor condensates, the spin dependence of the collisional interaction gives rise to spin-mixing of the internal states of the matter wave \cite{Ho98, Ohmi98, Law98}.  For a spin-1 condensate, the mixing of the 3 internal states is described by the interaction Hamiltonian:
\begin{equation}
\hat{H}_{s}= \hbar \chi (\hat{a}_{0}^{2}  \hat{a}_{+1}^{\dagger} \hat{a}_{-1}^{\dagger} + \hat{a}_{0}^{\dagger 2}  \hat{a}_{+1} \hat{a}_{-1})
\label{eq1}
\end{equation}

\noindent where $\hat{a}_{i}$ is the annihilation operator of the $i$th spin mode, $\hbar \chi$ is the spin interaction energy per particle and the spin modes are assumed to have the same spatial wavefunction \cite{Law98}. This Hamiltonian gives rise to spin-changing collisions between two $m_F =0$ atoms and a pair of $m_F = \pm1$ atoms, constrained by the conservation of angular momentum. With the identification of $\hat{a}_{0}$ as the pump mode and the $\hat{a}_{\pm1}$ modes as the signal and idler, this Hamiltonian is formally identical to the optical four-wave mixing Hamiltonian well-known in quantum optics \cite{Goldstein99}. If the $+1$ and $-1$ modes are initially unoccupied, subsequent parametric amplification of the vacuum states yields a two-mode squeezed state in which the two modes are exactly correlated in number. This process forms the basis for many proposals for generating atomic squeezing and entanglement using spin-1 BECs \cite{Pu00, Duan00, Duan02, You02, Sau10}.

Most spinor BEC experimental work has focused on the semi-classical or mean-field limit, but recently, in studies of the parametric amplification process, the effects of vacuum fluctuations were observed as excess (super-Poissonian) amplification noise \cite{Lett09b, Leslie09, Klempt10}. Thus far however, the experiments have not had the capability to detect sub-Poissonian quantum correlations necessary to demonstrate squeezing.

Here, we report on the measurement of sub-Poissonian spin correlations generated by four-wave spin mixing in a spin-1 BEC. The fluctuations in the relative numbers of atoms in the $+1$ and $-1$ modes are reduced by $-7$ dB ($-10$ dB corrected for detection noise) below the SQL following 4WSM from a pure $F=1$ $m_F =0$ condensate. This constitutes observation of relative number squeezing in the same sense as measurements of relative intensity squeezing observed in optical four-wave mixing experiments, although we stress that these observations are insufficient to prove quadrature squeezing or entanglement. Our results represent an almost 15-fold improvement in relative number squeezing observed in four-wave mixing in \cite{Jaskula10}, where the amount of detected squeezing was limited by the large number of outgoing collisional modes and the corresponding small number of atoms per mode.  Comparable levels of relative number squeezing were recently reported in the collisional de-excitation of a 1D condensate, where the reduced dimensionality restricted the number of available modes \cite{Schmied11}.

For our experiments, we utilize all-optical trapping techniques to create $^{87}$Rb condensates localized in a single anti-node of a CO$_2$ laser standing wave potential. A high magnetic field gradient is applied during the evaporation to create a pure $F = 1~m_F = 0$ condensate containing $3800$~atoms \cite{Chang04}. The trap frequencies are $(\omega_{\perp}, \omega_z) = 2\pi \times (340, 3500)$~Hz, which correspond to a peak condensate density of $n_0 = 6.5 \times 10^{14}$~atoms/cm$^3$, and Thomas-Fermi radii $(r_{\perp},r_z) = (3.2, 0.31)$~$\mu$m.

In order to observe quantum correlations, it is necessary to be able to measure atom numbers with a precision exceeding the corresponding Poissonian limit.  We employ spin state resolved fluorescence imaging using a large numerical aperture objective and a low-noise, high quantum efficiency CCD camera. The condensate is released from the trap and probed after a free expansion of 7.5~ms. During the first 4~ms of the expansion, a Stern-Gerlach magnetic field is applied to spatially separate the three spin components. The condensate is then probed for 100~$\mu$s with three pairs of orthogonal laser beams. The beams are detuned 5~MHz to the red of the $F = 2 \longleftrightarrow F' = 3$ cycling transition and have an intensity of 21 mW/cm$^2$. To avoid a decay out of the cycling transition, we simultaneously turn on a repump beam resonant with the $F = 1\longleftrightarrow F' = 2$ transition. The resulting fluorescence signal is collected by a CCD camera. The atom detection noise is limited by the technical noise of the CCD camera $\sigma_{cam}$, the photon shot noise of background scattered light $\sigma_{scatt}$, and the photon shot noise of the atom fluorescence $\sigma_{PSN}$. The first two noise sources are independent of the atom number and provide an overall detection noise floor $\sigma^2_{bkg} = \sigma^2_{cam}+\sigma^2_{scatt} = (8\, \text{atoms})^2$ per spin component.  The photon shot noise of the atom signal is $\sigma^2_{PSN} = N_{atoms}/a$, where $a$ is the mean number of detected photons per atom.

\begin{figure}
\centering
\includegraphics*[width=3.3in, height=3.00in, keepaspectratio=true]{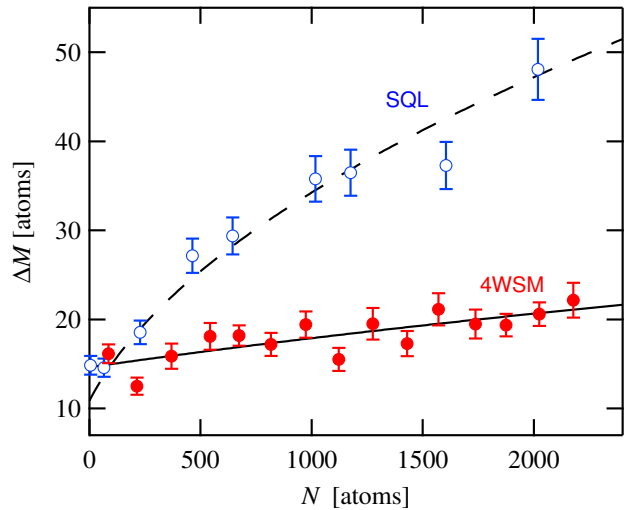}
	\caption[]{Fluctuations of the relative population of the $+1$ and $-1$ spin states, $\Delta M$, versus the number of atoms in the two states, $N$.  The open circles are for the CSS, and the closed circles are for  4WSM.  The error bars correspond to the statistical uncertainty in the measurement of the fluctuations given the finite number of measurements.  The dashed line is a curve fit to the CSS data, and the solid line is a curve fit to the 4WSM data.}
\label{Fig:DeltaM}
\end{figure}

We first evaluate the standard quantum limit (SQL) of our system using using the quantum projection noise of a coherent spin state (CSS) of the 3 Zeeman levels in the $F = 1$ manifold. The CSS is created by applying a radio frequency (RF) pulse to a pure $F =1, m_F =0$ condensate tuned to resonance with the $m_F = \pm1$ states, which are shifted in energy by an applied 430~mG magnetic field. The RF field induces Rabi flopping between the initial $m_F = 0$ state and the $m_F = \pm1$ states, whereby the $m_F = \pm1$ states are populated with equal probability.  Defining atom number operators for the population difference of the $+1$ and $-1$ modes, $\hat{M}  =  \hat{a}_{1}^{\dagger} \hat{a}_{1} -  \hat{a}_{-1}^{\dagger} \hat{a}_{-1}$ and the total population $\hat{N} = \hat{a}_{1}^{\dagger} \hat{a}_{1} +  \hat{a}_{-1}^{\dagger} \hat{a}_{-1}$, the CSS we create satisfies $\langle M\rangle = 0$ and exhibits Poissonian quantum projection noise, or ``shot-noise,'' $\Delta M_{QPN} = \sqrt{N}$.  To establish the SQL, we measure the population difference of the $+1$ and $-1$ spin states of the CSS as a function of $N$ by varying the length of the RF-pulse.  Each measurement is repeated 100 times in order to acquire sufficient statistics, and from these measurements, the standard deviation of the population difference $\Delta M$ is determined. The results are shown in Fig.~\ref{Fig:DeltaM}, together with a fit to a curve incorporating the atom detection noise according to: $(\Delta M)^2 = \sigma^2_{bkg} + \sigma^2_{PSN}+ (\Delta M_{QPN})^2$. The data show the expected $\sqrt{N}$ behavior for $\Delta M\gtrsim20$, and the parameters determined from the fit, $a = 18.2 \pm 0.9$ photons/atoms and  $\sigma_{bkg} = 11 \pm 4$ atoms, are consistent with the estimation of the fluorescence scattering rate, the optical collection efficiency and the measured detection noise floor.

\begin{figure}
\centering
\includegraphics*[width=3.29in, height=3.30in, keepaspectratio=true]{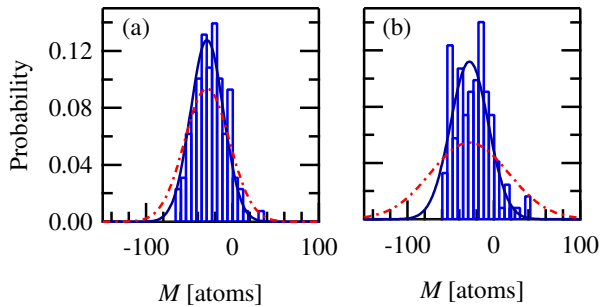}
	\caption[]{Histograms of the population difference $M$ after 4WSM for a) $\langle N \rangle = 674$ atoms and b) $\langle N \rangle = 2025$, where $\langle N \rangle$ is the average total number of atoms in the $m_F = \pm1$ states for the corresponding bin. The solid lines are Gaussian fits to the distributions, with corresponding widths of $19 \pm 1$ and $22 \pm 1$ atoms, respectively. Both these values are lower than the corresponding SQL with width $\sqrt{\langle N \rangle}$ ($=26$ and $=45$~atoms, respectively), indicated with dashed lines.}
\label{Fig:SpinMixHist}
\end{figure}


We now turn to the study of spin correlations dynamically generated by 4WSM in the condensate. The condensate is created in a magnetic field $B \approx 2$~G, such that the Zeeman energy dominates the spinor interaction and the condensate remains in the $m_F = 0$ state . Spin mixing is induced by rapidly lowering the magnetic field to 360~mG in 10~ms, at which point the 4WSM interaction dominates over the Zeeman energy \cite{Chang04}. The condensate is allowed to dynamically evolve for 200~ms, during which the non-linear spin interaction generates quantum spin correlations. The spin states of the condensate are then measured using the same procedure as above. The experiment is repeated 1300 times in order to acquire data sets for different degrees of spin-mixing. This data is then binned according to $N$ with a bin size of $150$ atoms such that each data set contains a sufficiently large number of trials (typically $> 60$) to determine the fluctuations of the population difference, $\Delta M(N)$. Two such data sets are shown as histogram plots in Fig.~\ref{Fig:SpinMixHist}. We compare these data sets to the Poissonian distribution for the corresponding CSS  with the same mean atom number (dashed line). In both cases, the measured fluctuations, fit to a Gaussian curve (solid line), are smaller than the CSS, demonstrating sub-Poissonian fluctuations. The fluctuations of the population difference for 4WSM are compared with the corresponding CSS measurements for values of $N$ up to 2200 atoms in Fig.~\ref{Fig:DeltaM}.

\begin{figure}
\centering
\includegraphics*[width=3.3in, height=3.00in, keepaspectratio=true]{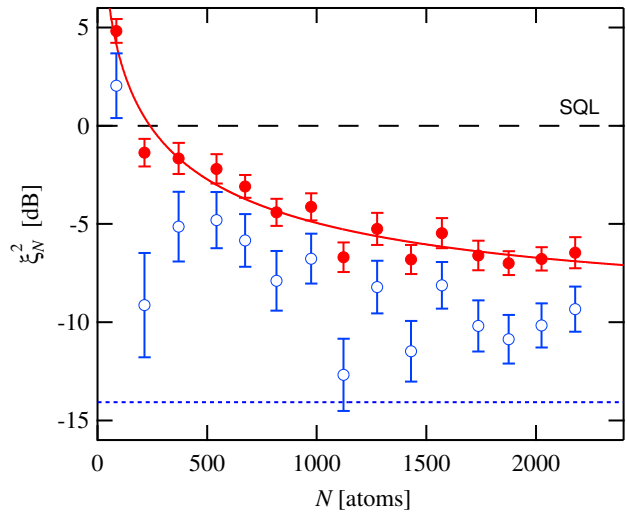}
	\caption[]{Relative number squeezing versus the degree of spin-mixing. The solid circles are the raw data and the open circles are corrected for detection noise. The error bars include the uncertainty due to the finite size of the sample and the uncertainties in the calibration constant and background scatter. The  solid line is a fit to the theoretical prediction including the effects of detection noise and atom loss. The dotted line shows the squeezing limit due to only atom loss, and the dashed line indicates the SQL.
}
\label{Fig:RelNumSqu}
\end{figure}


To quantify how much the relative atom fluctuations are suppressed compared to Poissonian statistics, we define the relative number squeezing parameter $\xi^2_N =\Delta M^2/N$.  This is plotted in Fig.~\ref{Fig:RelNumSqu} versus the average atom number. Even for a small number of $m_F = \pm1$ generated atoms ($N > 214$), the relative number squeezing parameter is under the SQL ($\xi^2_N<0$). A maximum squeezing of $-7$~dB is directly measured for large populations of the $\pm1$ spin states, $N>1500$.  Using the atom detection noise that was measured above, it is possible to infer a `corrected' squeezing parameter that would be obtained with detection improvements.  This is also plotted in Fig.~\ref{Fig:RelNumSqu} for comparison. A maximum relative number squeezing exceeding $-10$ dB is inferred from these data.

The maximum squeezing that we observe is limited by both the atomic detection noise, which is a technical limitation, and by atomic loss processes, which limits the maximum squeezing that is created in the condensate.  The finite lifetime of the condensate ($\tau = 1.5$~s) due to collisions with thermal atoms in the residual vacuum and three-body collisional losses gives rise to uncorrelated atomic loss.  These losses  degrade the spin correlations by an amount $\sigma^2_{\text{loss}} = p(1-p)N^0 = pN$, where $N^0$ is the number of atoms in the $m_F = \pm1$ states without loss, and $p$ is the probability that an atom in the $m_F = \pm1$ states was lost. Combining the atomic losses with the detection noise leads to an overall noise floor $\Delta M$ given by $(\Delta M)^2  = \sigma^2_{\text{bkg}} + \sigma^2_{\text{PSN}}  + pN$.  This is used to fit the 4WSM data using the value $\sigma_{PSN}$ from above, and the results of the fit are shown as solid curves in Fig.~1 and Fig.~3.  The best fit indicates a background noise of $\sigma^2_{\text{bkg}} = 15 \pm 1$~atoms, which is slightly larger than for the CSS data but within the margin of error and the day to day variations of the background. The fit also yields a loss probability of $p = 0.05 \pm 0.02$, which is consistent with the 4\% value determined from the temporal evolution of the 4WSM process discussed below. The agreement of the data with the curve fit is satisfactory and shows that the maximum observed squeezing is quantitatively consistent with the detection and loss limits.

\begin{figure}
\centering
\includegraphics*[width=3.3in, height=5.00in, keepaspectratio=true]{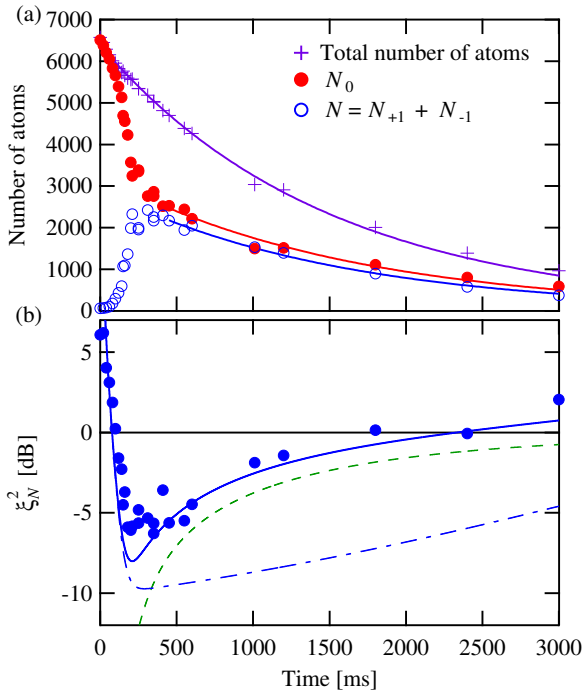}
\caption[]{(a) The number of atoms in the condensate and the spin populations versus 4WSM time. The curves are exponential fits to the data. (b) Relative number squeezing versus  time. The circles are the raw (uncorrected) data.  The dash-dotted line indicates the detection noise limit and the dashed line indicates the atom loss limit.  The solid line includes both limits. }
\label{Fig:AtomLossdB2}
\end{figure}

The dynamical evolution of the relative number squeezing is shown in Fig.~\ref{Fig:AtomLossdB2}, together with the population of the atomic spin states and the total number of atoms in the condensate.  As is evident from the figure, the occupation of the $m_F = \pm1$ modes grows rapidly due to spontaneous 4WSM and reaches a 50\% steady-state population relative to the total number of atoms after 250~ms. For later times, the relative spin populations remain constant, indicating uncorrelated losses, while the populations of the individual spin states decay with the same time constant ($\tau = 1.5$~s) as the overall decay of the condensate.  For these data, condensates in two optical lattice sites were used so the total number of atoms was larger. The evolution of the relative number squeezing shown in Fig.~\ref{Fig:AtomLossdB2} reveals explicitly the relative importance of detection noise and atom loss as a function of time.  Initially, when the number of atoms in the $m_F = \pm1$ states is very small, the observed relative number squeezing is dominated by the atom detection noise floor (dashed-dotted line). At later times, atomic loss becomes the predominate limitation to squeezing (dashed line). The solid line includes both limiting factors and agrees well with the observed data.

In summary, we have observed sub-Poissonian fluctuations in the relative atom numbers create by four-wave spin mixing. The fluctuations are reduced up to 7~dB,  which is limited by the atom loss out of the trap during the 200~ms of spin-mixing. This is the first demonstration of sub-Poissonian spin statistics in a spin-1 condensate and provides a solid foundation for future experiments involving the demonstration of two-mode quadrature squeezing and entanglement in a spinor condensate \cite{Duan02, You02, Sau10}.  Indeed, we have recently demonstrated a large degree of quadrature squeezing in this system that we will discuss in a separate report.

This work was supported by the NSF Grant No. PHYS-0605049. We gratefully acknowledge Li You for helpful conversations.



\bibliography{RelNumSqu}

\end{document}